# Spectral Bandwidth Recovery of Optical Coherence Tomography Images using Deep Learning

**Timothy T. Yu[1,4], Da Ma[2,1*], Jayden Cole[1], Myeong Jin Ju[3,4], Mirza F. Beg[1], and Marinko V. Sarunic[1,5,6*]**

[1]*Engineering Science, Simon Fraser University, Burnaby BC V5A1S6, Canada*
[2]*Wake Forest University School of Medicine, Winston-Salem, NC, 27151, USA*
[3]*Dept. of Ophthalmology and Visual Sciences, University of British Columbia, Vancouver, BC, V5Z 3N9, Canada*
[4]*School of Biomedical Engineering, University of British Columbia, Vancouver, BC, V5Z 3N9, Canada*
[5]*School of Biomedical Engineering, University of British Columbia, Vancouver, BC, V5Z 3N9, Canada*
[5]*Institute of Ophthalmology, University College London, London, UK*
[6]*Department of Medical Physics and Biomedical Engineering, University College London, United Kingdom*

**dma@wakehealth.edu*
*[*]m.sarunic@ucl.ac.uk*

## 1. Abstract

Optical coherence tomography (OCT) captures cross-sectional data and is used for the screening, monitoring, and treatment planning of retinal diseases. Technological developments to increase the speed of acquisition often results in systems with a narrower spectral bandwidth, and hence a lower axial resolution. Traditionally, image-processing-based techniques have been utilized to reconstruct subsampled OCT data and more recently, deep-learning-based methods have been explored. In this study, we simulate reduced axial scan (A-scan) resolution by Gaussian windowing in the spectral domain and investigate the use of a learning-based approach for image feature reconstruction. In anticipation of the reduced resolution that accompanies wide-field OCT systems, we build upon super-resolution techniques to explore methods to better aid clinicians in their decision-making to improve patient outcomes, by reconstructing lost features using a pixel-to-pixel approach with an altered super-resolution generative adversarial network (SRGAN) architecture.

## 2. Introduction

Optical coherence tomography (OCT) is a non-invasive imaging modality that allows for high-resolution volumetric visualization of the retina, the light-sensitive tissue at the back of the eye. OCT is the gold standard diagnostic for diseases such as age-related macular degeneration (AMD) and diabetic macular edema (DME), but is not a widely accepted modality for diabetic retinopathy (DR) diagnosis and monitoring due to the limited field of view (FOV) [1]. OCT facilitates the characterization of retinal thickness changes and abnormalities that are indicative of DR which if integrated as a secondary diagnostic modality, may benefit the outcome of patients with DR. We have previously demonstrated that the vasculature outside the parafovea contains features indicative of early changes from DR [2]. Hence, technology is advancing towards wide-field OCT systems to capture more details of the retina.

As the OCT hardware advancements move towards capturing a wider FOV on the retina, there is an effort to minimize motion artefacts and patient discomfort that often accompany the longer acquisition from an increased FOV. Increasing the speed of the acquisition system often results in engineering compromises that reduce the spectral bandwidth of the OCT system, and hence lower the axial resolution. Methods like montaging [3] and motion-tracking software [4] are some techniques that have been explored to minimize the negative implications of wide-field OCT systems. In addition, machine learning has been explored for feature reconstruction of OCT-angiography (OCTA) to improve the image quality for clinical utility [5].

In this study, we simulate the narrower spectral bandwidth on OCT volumes and investigate the use of a generative adversarial network (GAN). For image-to-image generation, many implementations use a pixel-to-pixel (pix2pix) GAN, which has the encoder/decoder in the generator. We used a pix2pix approach leveraging a modified super-resolution GAN (SRGAN) [6] architecture to recover high-resolution features in the OCT B-scans. The SRGAN is comprised of a VGG-19 style discriminator and a generator with residual blocks and subpixel convolutional layers. We have modified the SRGAN architecture by deepening the discriminator and generator and removed the pixel up-sampling layers from the generator for our pix2pix implementation.

We were limited by the size of our dataset and leveraged transfer learning from an open-source natural flower dataset [7], referencing a similar SRGAN approach used to upscale radiographs [8], to facilitate the convergence of our deep neural networks (DNNs). The flower dataset contains images with dense and well-defined edges from complex features of different flowers, petals, and seeds. These features exaggerate the blurring process and may prove useful when used as initialized weights for our OCT-SRGAN.

The original architecture was designed to upscale images and reconstruct features lost from downsizing. Rather than utilizing the OCT-SRGAN to reconstruct a larger image from a resized smaller image, we investigate the use of an OCT-SRGAN to reconstruct reduced spectral resolution in the axial direction (A-scan blurring) through transfer learning and demonstrate the ability to reconstruct lost features using learning-based approaches. Our findings suggest that DNNs may benefit clinicians if developed in parallel with wide-field OCT systems by reconstructing features lost due to the reduced narrower spectral bandwidth that often accompanies OCT systems with faster A-scan line rates. Our main contribution is to adapt the feature recovery using an SRGAN for the super-resolution of OCT B-scans. We present our preliminary results while we continue to develop our OCT-SRGAN and will branch into alternative learning-based methods like feature pyramid networks. This experiment was done in a two-step approach as described in Figure 1: (1) preliminary dataset consisting of 16 eyes to understand the impact of transfer learning; (2) utilize the full dataset of 35 eyes to compare the performance of reconstruction in the spatial versus spectral domain.

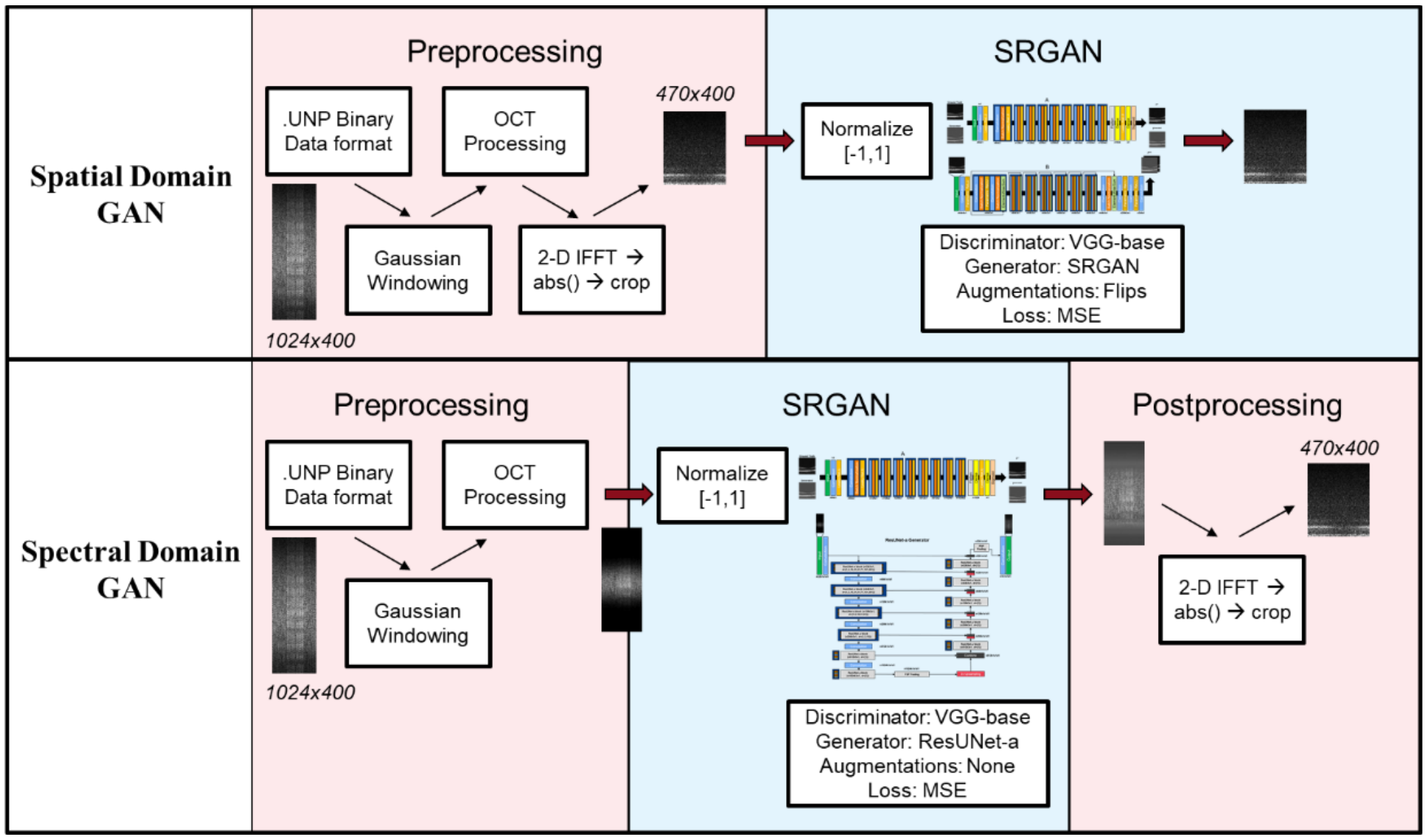

Figure 1 Parallel schematics comparing the spatial and spectral super-resolution GAN processing pipeline.

## 3. Methods

### *OCT dataset and preparation*

This study included 35 eyes (including pathological) 27 unique patients, each imaged on a ~2x2mm FOV centered on the fovea using a 1060nm swept source OCT prototype adaptive optics system. We utilized B-scans that visualize the retinal cross-section. An initial subset of this dataset consisted of 16 eyes was initially used to validate transfer learning from natural images. Each B-scan was cropped from 1024x400 to 240x400 (axial x lateral position) for the preliminary experiment and were cropped to 470x400 and processed as strips of 20 A-scans for the spatial domain portion of the second experiment. The image dimensions for the spectral domain GAN were unchanged before evaluation. The full dataset of 35 eyes consisted of 14,000 B-scan samples or 5,600,000 independent A-scan samples. The dataset was split into 60%, 20%, 20% for training, validation, and testing, respectively. 21 eyes were used for training (8,400 B-scans or 3,360,000 A-scans), 7 eyes for validation (2,800 B-scans or 1,120,000 A-scans), and 7 eyes (2,800 B-scans or 1,120,000 A-scans) were allocated for testing. Adjacent B-scans contain similar information. Thus, every 8 B-scan was used to allow enough spacing between acquisitions to minimize the chance of overfitting to repeating consecutive scans. The images were shuffled within the training set prior to training. Special attention was made to ensure that eyes from the same patient was used for either training, validation, or testing. Since the flower images were 3-channel RGB images, when training the neural networks on the flower images, we selected the first channel to form a 1-channel input to replicate the domain of OCT images.

The axial resolution in OCT images is related to the spectral characteristics of the light source. A commonly used expression for the axial resolution is the coherence length, lc, given by $l_c = \lambda_o^2/\Delta\lambda$, where $\lambda_o$ is the central wavelength, and $\Delta\lambda$ is the spectral bandwidth. By using our swept source OCT prototype system, we had access to data in the wavelength domain and performed Gaussian windowing on the spectrum and hence reduced the axial resolution. We utilized the MATLAB function gausswin with $\alpha = 8$, *chosen based on earlier trials of training and the appearance of the OCT B-scan image output, where the coefficients of a gaussian window (w) are as described in*[1]. This gaussian windowing mask was element-wise multiplied with the spectral domain OCT data, as shown in Figure 2.

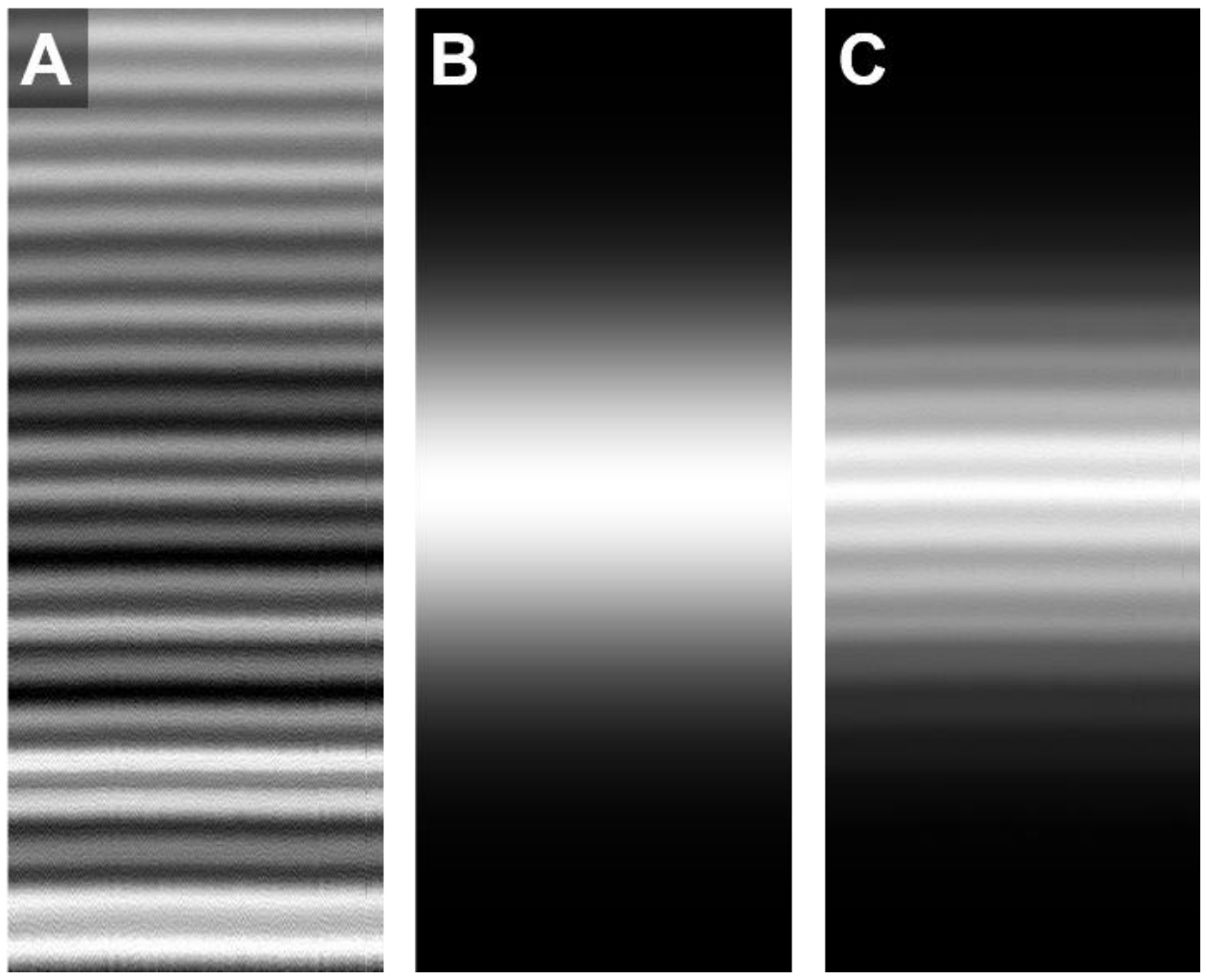

Figure 2 Windowing process of our OCT data in the spectral domain (A). A gaussian windowing mask (B) is used to reduce the axial resolution with α = 4 to generate spectral domain gaussian windowed data (C).

## Flower Dataset and Preparation

We used an open-sourced flower dataset compiled by the team at TensorFlow [7] to initialize the weights of our OCT-SRGAN. The flower dataset is comprised of 3,670 flower images. The black borders surrounding the floral images were removed and the images were reshaped to 240x400 and randomly shuffled before being allocated for training (3303) and testing (367).

The artefacts introduced to the flower dataset should mimic the appearance of the reduced axial resolution OCT B-scans. Hence, we generated the low-resolution images by convolving the original high-resolution (HR) inputs with a 1xn mean filter to smoothen each pixel vertically with n values of 3, 5, 7, 9, and 11 pixels, as shown in Figure 3.

[1] https://www.mathworks.com/help/signal/ref/gausswin.html

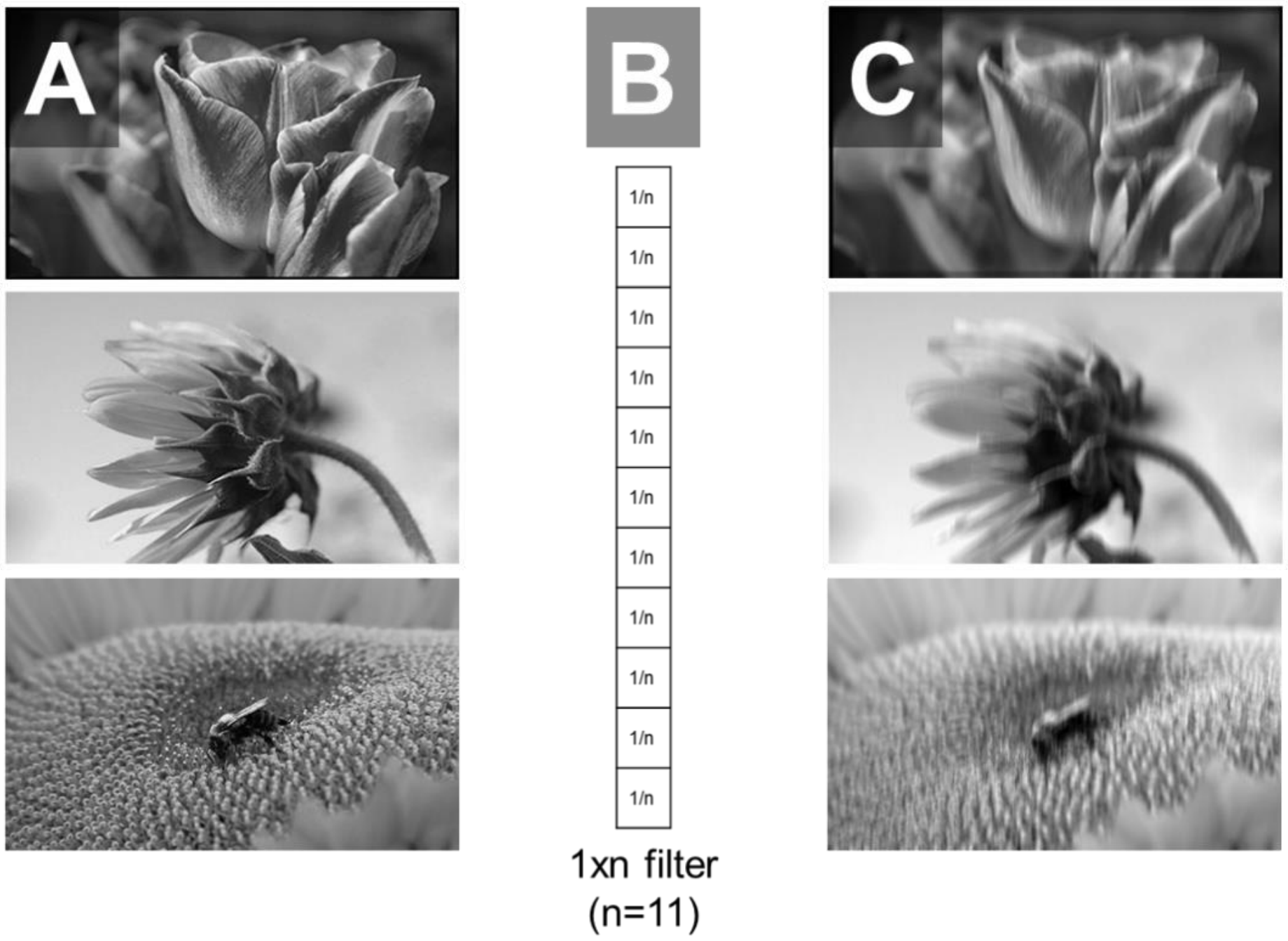

Figure 3. Graphical representation of the windowing process of high-resolution flower images (A) using a vertical 1xn, in this case n=11, mean filter (B) to generate mono-directional (vertical) smoothened images (C).

## Data Augmentation and GAN Training Techniques

In reconstructing in the spatial domain, data augmentation was achieved through horizontal and vertical flip using the ImgAug library to increase the effective dataset size and improve the generalizability. Since the data experienced smoothening in A-scans (vertical direction) only, rotations were not considered. Additionally, random noise was not introduced through augmentation to not interfere with the reconstruction of the speckle pattern in OCT B-scans. In both domains, the data was augmented during preprocessing and the did not undergo further augmentations during training. These augmentations performed on the spectral domain fringe data included a random center and width of the Gaussian windowing filter, as demonstrated in Figure 4.

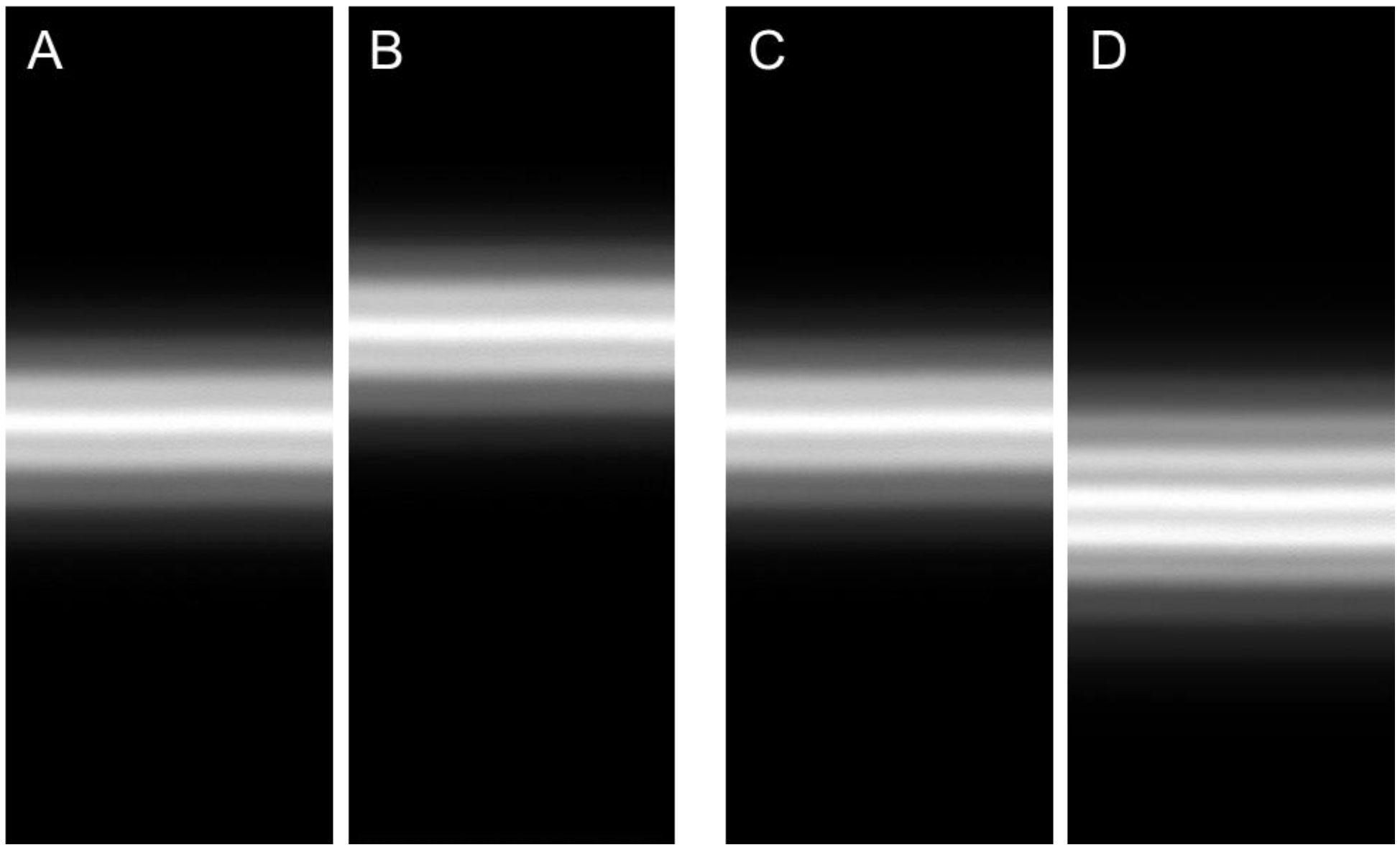

Figure 4 An illustration of the random augmentations performed during the preprocessing of the spectral domain fringe data: (A & C) original Gaussian windowed fringe data; (B) center-shifted; (D) center-shifted and widened.

We refer to the previous literature [9–13] for neural network training procedures to improve our GAN framework and encourage convergence. Soft labels are used in classification neural networks to decrease the error rate and have been adapted to the adversarial component of a GAN [9–11]. Instance noise is a technique where the discriminator's training labels are randomly flipped [12], [13]. Both techniques were implemented to improve the stability of GAN training by reducing the ambiguity between the generated and ground truth samples which promotes training convergence. We implemented soft labelling by randomly distributing real labels between (0-0.1) and generated labels from (0.9-1). In addition, instance noise was introduced by randomly, with a 5 percent chance, providing the discriminator with an incorrect label. The discriminator was trained on an entire batch of the real ground truth data and followed by a batch of the generated data. This minibatch feature approach allowed the discriminator to compare an example of a minibatch of generated samples to the real samples and allowed the discriminator to detect similarities across the minibatches [9], [14].

## Neural Network Design

### Spatial-domain Feature Recovery Network

In the first step, we propose a variation of the SRGAN proposed in the Literature [6]. Instead of to upscales the smaller image to a larger image through subpixel convolutional layers (PixelShuffler x2), our implementation of the SRGAN removes the subpixel resolution layers and instead, utilizes the SRGAN architecture for a pix2pix application to preserve the B-scan resolution, as shown in Figure 5.

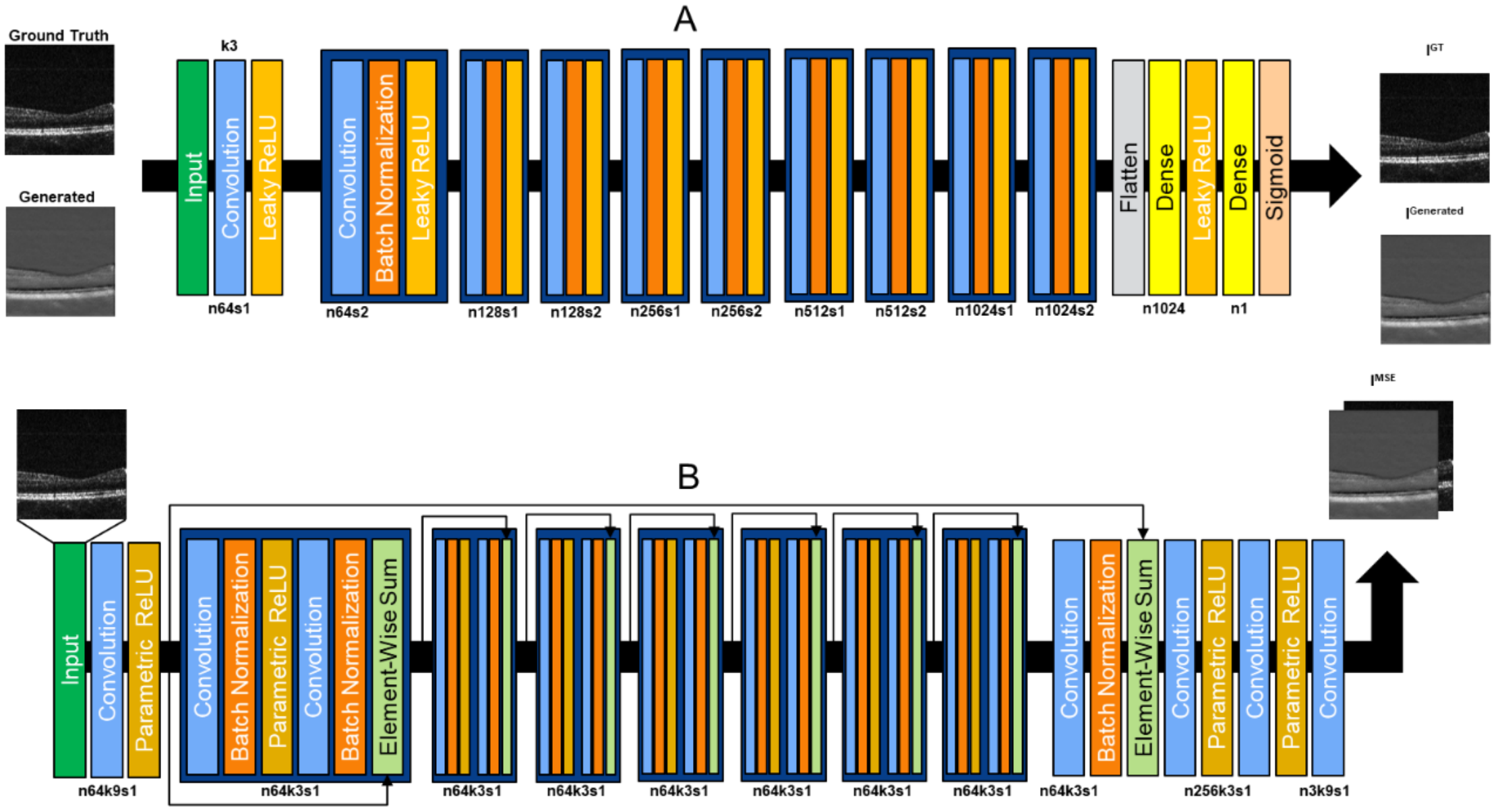

Figure 5 Super-resolution Generative Adversarial Network (SRGAN) architecture. Architecture (A) is the discriminator based off VGG networks and architecture (B) is the generator comprised of residual of blocks. The loss functions are the discriminator ground truth (lGT) and generated image (lGenerated) binary cross entropy loss and generated/ground truth mean squared error (lMSE).

Compared to the SRGAN [6], we used a deeper discriminator and generator because deeper networks have been shown to yield better results with the tradeoff of being more difficult to train [15]. The discriminator consisted of convolutional blocks that contain 2D convolutional layers that utilizes more filters in deeper layers followed by batch normalization and a Leaky Rectified Linear Unit (LeakyReLU). Batch normalization has been found to improve the optimization of GANs [9], [16] and LeakyReLU has been found to generate better results, especially for higher resolution implementations [16]. In the generator, we used a parametric ReLU which is a LeakyReLU with a learnable negative slope along with batch normalization and skip connections through element-wise addition to form a residual block. Skip connections allow the network to pass forward simple features that may be difficult to learn through convolutional filters [6], [17].

## Spectral-Domain Feature Recovery Network

In the second step of the study comparing super-resolution in the spatial versus spectral domain, all the components within the neural network including convolutional filters, pooling filters, among others, utilized vertical 1-D filters. This ensured that the neural network would explicitly process the A-scans independently. The method of Gaussian windowing was only performed on A-scans, and each was independent from the others.

The spectral domain reconstruction processed of images with relevant information near the center of the B-scan with Gaussian-distributed diminishing intensity towards the end of the image. The generator was updated to account for the distance between pixels using dilation rates in the convolutional layer, which increase the spacing of the convolutional filters. Referencing the published ResUnet-a architecture [18], a wide range of dilation rates were leveraged in parallel where each block of batch normalization and activation layers were followed by a 1D convolutional layer with different dilation rates. Upwards of 8 parallel blocks were leveraged, each with different dilation rates, to allow the neural network to explore the relationship between pixels that were higher in distance. The parallel convolutional blocks with different dilation rates are combined through addition with an identity function through skip connections. Figure 6 graphically outlines the different building blocks used including the application of dilation rates in parallel to construct the ResUNet-a generator, as shown in Figure 7.

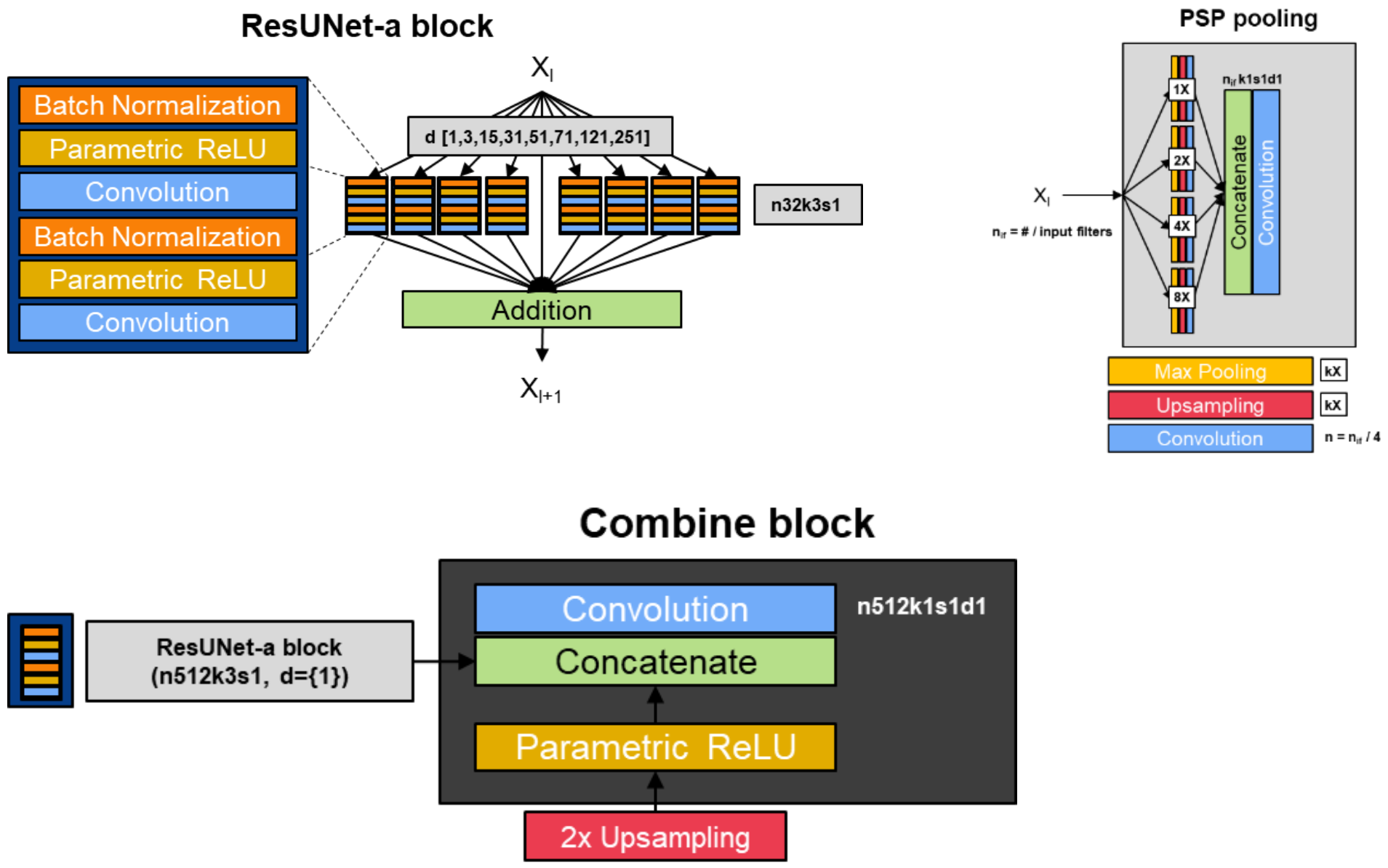

Figure 6 ResUNet-a building blocks as described [18].

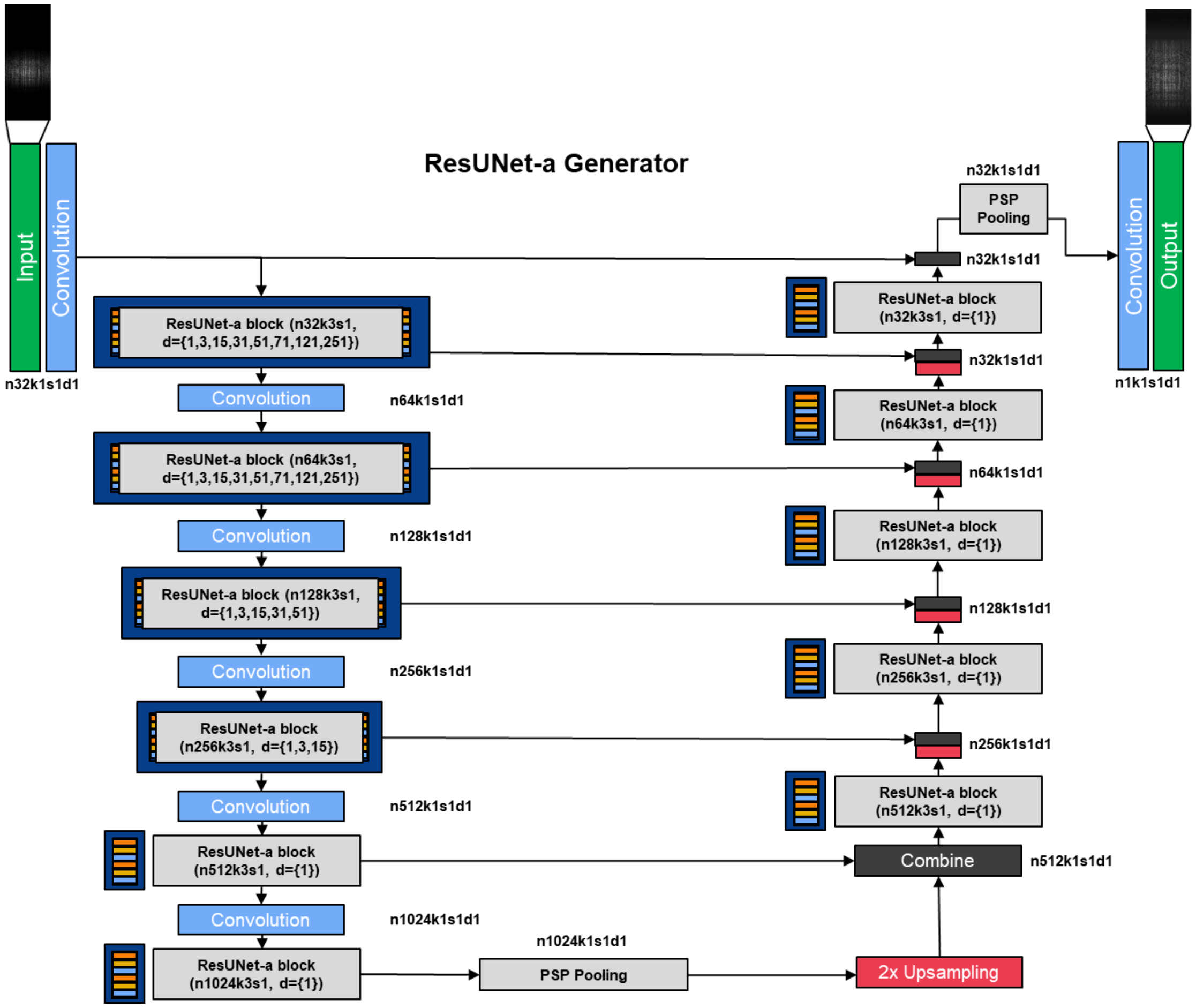

Figure 7 ResUNet-a architecture used for reconstructing in the spectral domain.

## Experimental Settings for Training Feature Recovery Network

DNN training for both experiments were implemented in a similar manner. The optimizer used was Adam with β1 = 0.5. The optimal learning rate was found through quick training sessions with different learning rate schedules. For all experiments, the optimal initial learning rates of 1x10-4 were used in both the discriminator and generator. For the preliminary experiment, the SRGAN was trained for 200 epochs with a batch size of 32. The second experiment was performed on images of size 1024x400 and was trained on 100 epochs with a batch size of 10. The SRGAN was trained and evaluated within 24 hours and 52 hours for the preliminary experiment and second experiment, respectively.

The content loss (lMSE) was calculated through pixel-wise mean squared error (MSE). The discriminator was trained separately on minibatches of ground truth and generated images and resulted in two binary cross-entropy losses (lGT, lgenerated) which refers to the generator's ability to fool the discriminator. Models that improved with a lower generator lMSE loss, higher discriminator lGT, or higher lgenerated loss, were saved. The preliminary experiment leveraged the SRGAN trained on the natural flower dataset that was used to initialize the weights for OCT B-scan training; all the neural network parameters were trainable. All DNNs were developed and evaluated in TensorFlow and the Keras API using Python 3.6.3 on Canadian supercomputer "Cedar" nodes powered by the NVIDIA Tesla V100-SXM2 GPU and 32GB RAM.

## Evaluation

The evaluation of GANs is often qualitative as there are minor artefacts that may drastically impact evaluation metrics like MSE, peak signal-to-noise ratio (PSNR), and structural similarity (SSIM) [19] among others. For example, in our implementation of reconstructing features, we may weigh the exact intensity of an image less than the sharpness of an OCT layer boundary and some evaluation metrics may weigh according to the neural network's function. Hence, one approach is to enlist human annotators to grade the performance of the GAN, also called the mean opinion score (MOS), based on the function and objective of the study [8], [9].

To improve the qualitative evaluation of the generated outputs, we cropped regions of the images near the retinal layers and enlarged the image using nearest-neighbour interpolation to preserve the pixel resolution and qualitatively compared to the ground truth and windowed/mean filtered images to better visualize and compare features. The models were quantitatively evaluated using MSE, PSNR, and SSIM across the entire test set for the entire image of the preliminary experiment. RMSE and SSIM were leveraged to evaluate the experiment on the full dataset.

## 4. Results – Preliminary Transfer Learning

A range of filter sizes (1x3, 1x5, 1x7, 1x11) was utilized to introduce vertical blurring to the flower neural network. The SRGAN trained on the 1x11 smoothened flower images yielded the best results when used as initialized weights for OCT B-scans. Therefore, in the transfer learning experiment, the neural network was initialized on the 1x11 smoothened flower image.

### *Qualitative Evaluation*

Our direct implementation of the OCT-SRGAN to reconstruct high-axial-resolution from low-axial-resolution B-scans successfully sharpened the features of the OCT image. As shown in Figure 9 (A - C), the same region is cropped and enlarged using nearest-neighbor interpolation for qualitative evaluation.

The results of our flower SRGAN and the performance of reconstructing vertical smoothening using a 1x11 mean filter, as shown in Figure 8, shows promising utility in recovering features lost from smoothening.

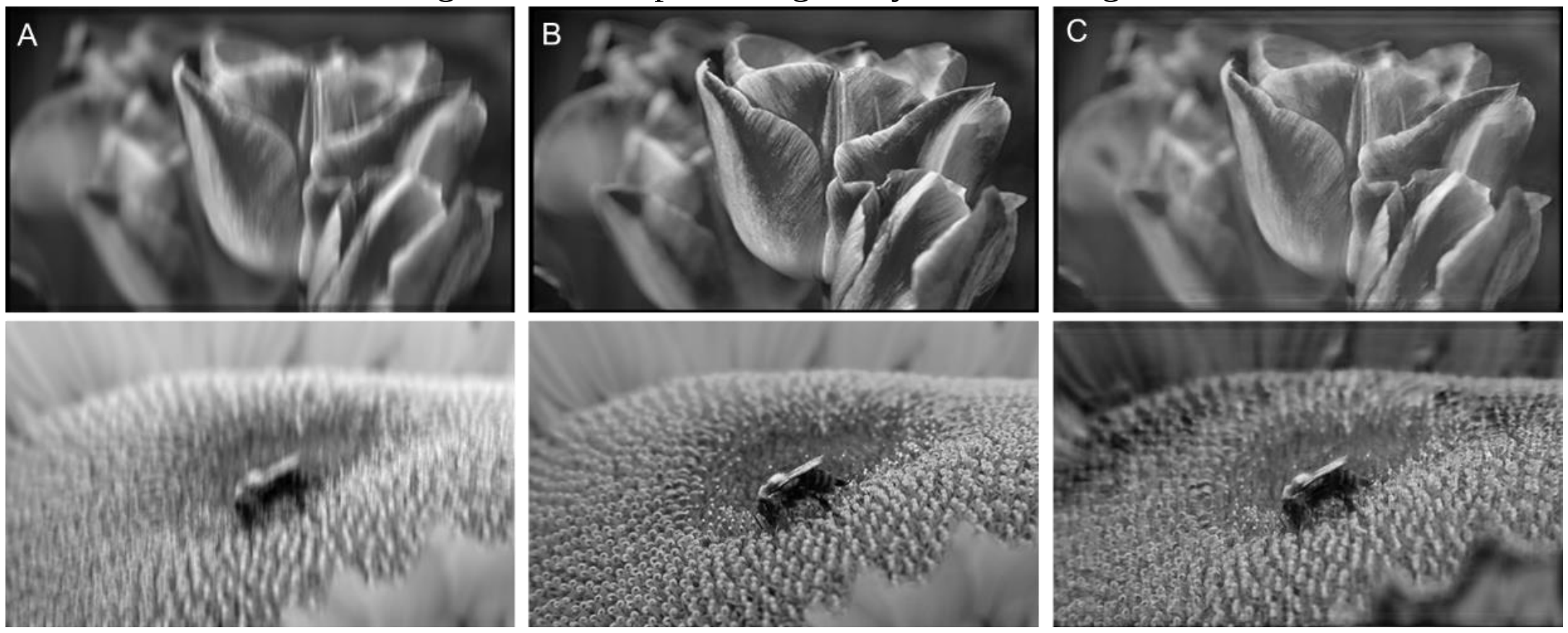

Figure 8 Flower SRGAN comparison between (A) 1x11 mean filtered, (B) ground truth, and (C) generated images from the test set.

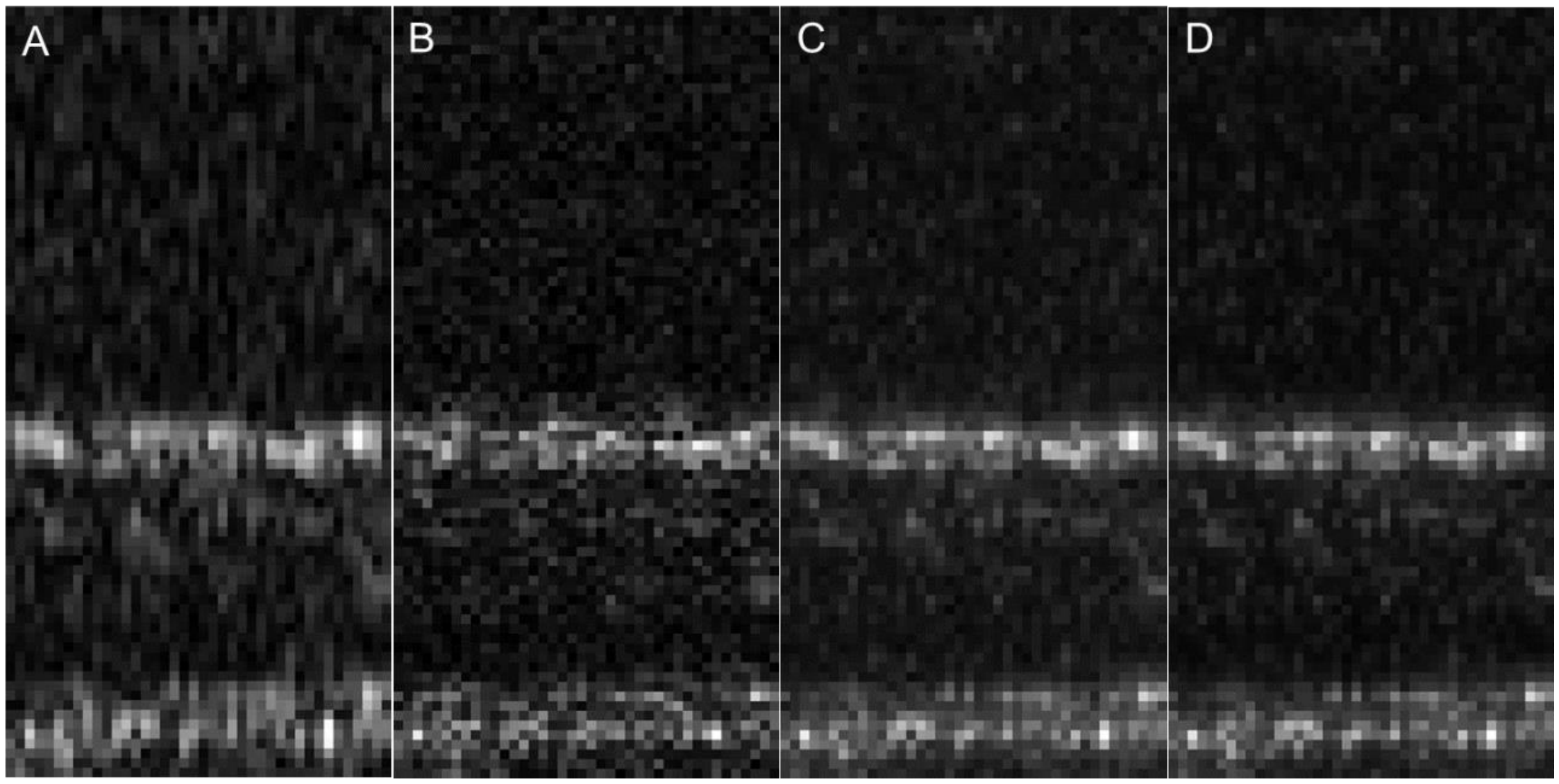

Figure 9 Results of SRGAN for OCT B-scan images. Samples from the generated images from the test set. Comparison between (A) windowed, (B) ground truth, and (C) generated B-scans without transfer learning, and (D) generated B-scans with transfer learning.

### Quantitative Evaluation

The whole OCT B-scan image was evaluated for similarities across the test set post-contrast adjustment between the generated and ground truth image using MSE and SSIM, as shown in Table 1. This was compared to the baseline comparison of the Gaussian windowed/vertically smoothened and ground truth images.

While it is difficult to quantitatively evaluate the performance of a generated output as the function or goal is often subjective, by standardizing the image and enhancing the contrast of all images using the same pipeline, the issues of differences in intensities were minimized. By doing so, we effectively catered our metrics to weigh the features more heavily than the absolute intensity. Table 1 shows that across the evaluation metrics, the generated B-scans more resembled the original high-resolution B-scans.

Table 1. Quantitative evaluation of the generated images (GEN) comparing the OCT-only with the natural images transfer learning approach. Mean values are shown with standard deviation in parentheses. This portion of the experiment was performed on a subset (16 eyes) of the entire dataset (35 eyes).

| Test | GEN vs. GT | | Windowed vs. GT | |
|---|---|---|---|---|
| | ***MSE*** | ***SSIM*** | ***MSE*** | ***SSIM*** |
| No Transfer Learning (16 Volumes) | **71.04 (36.67)** | **0.753 (0.059)** | 100.42 (38.66) | 0.713 (0.051) |
| With Transfer Learning (16 Volumes) | **68.07 (40.38)** | **0.767 (0.067)** | 100.42 (38.66) | 0.713 (0.051) |

**Best value** between generated and windowed for the specific test is bolded.

Mean squared error (MSE) and structural similarity (SSIM).

Ground truth (GT)

## 5. Results – Reconstruction in the Spatial Versus Spectral Domain

The method of preprocessing significantly affected the comparison between the training on the preliminary dataset compared to the full dataset since a different cropping algorithm was utilize. This section leverages the entire 1024x400 B-scan; conversely, the preliminary experiment cropped the images into 470x400 B-scans. However, within this experiment, the method of evaluation was consistent between both the spatial and spectral domain with minor differences in functions used for some final processing and evaluation steps.

### *Qualitative Evaluation*

For the spatial SRGAN, the full set of 35 eyes from 27 unique patients were used without transfer learning and cropped regions of two scans from the test set capturing the retinal layers are shown in Figure 10.

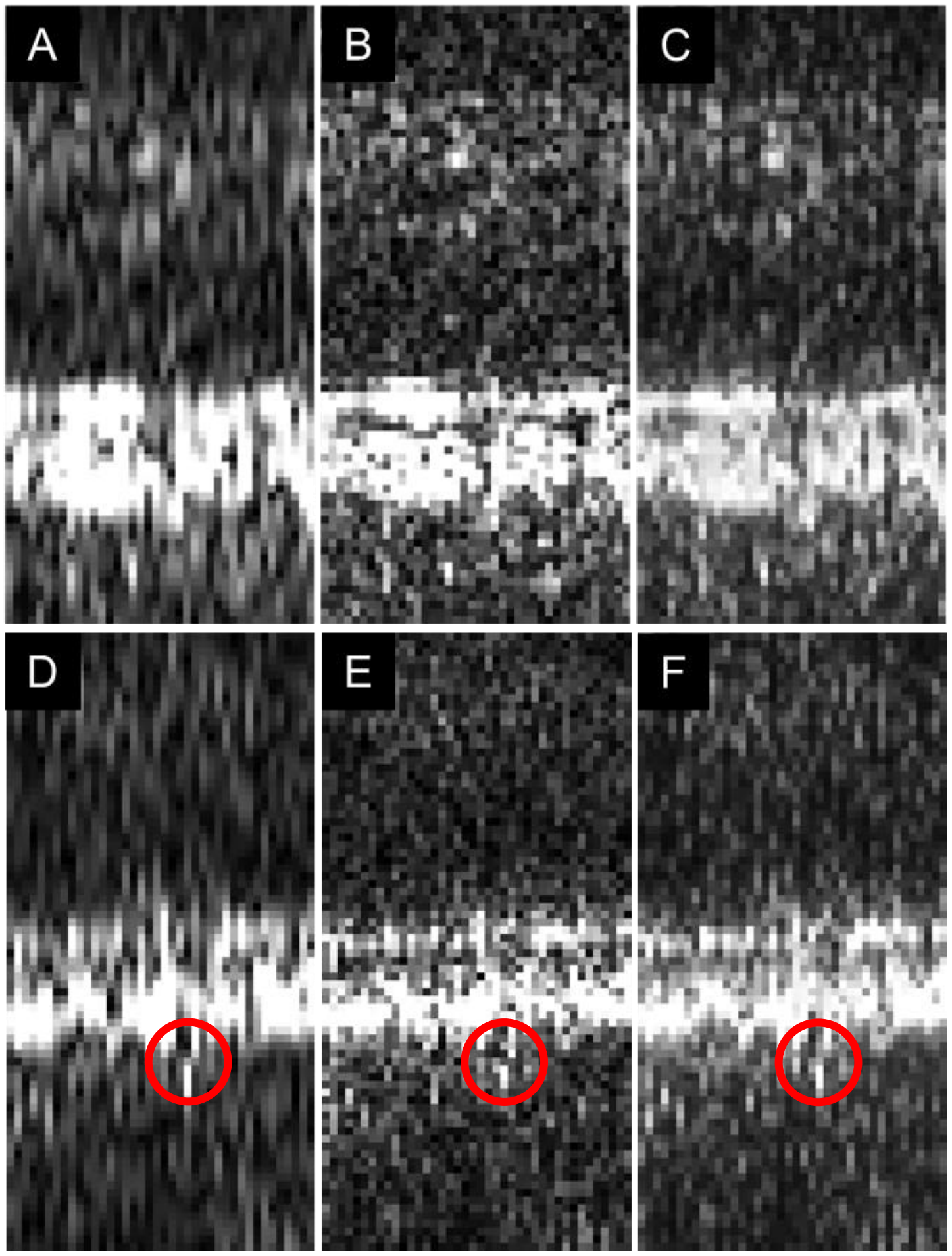

Figure 10. Results of spatial SRGAN without transfer learning on the full OCT B-scan dataset (35 eyes). Samples from the generated images from the test set. Comparison between (A & D) windowed, (B & E) ground truth, and (C & F) generated B-scans. The red circles highlight a region of interest.

The full dataset reconstructed from the spectral domain was used to train a model and a cropped region of an eye from the test set is shown in Figure 11. Another example from the test set can be seen in the Figure 12. The data is independent across A-scans and should be further explored as so. Figure 13 graphically compares the intensity of the fringe data.

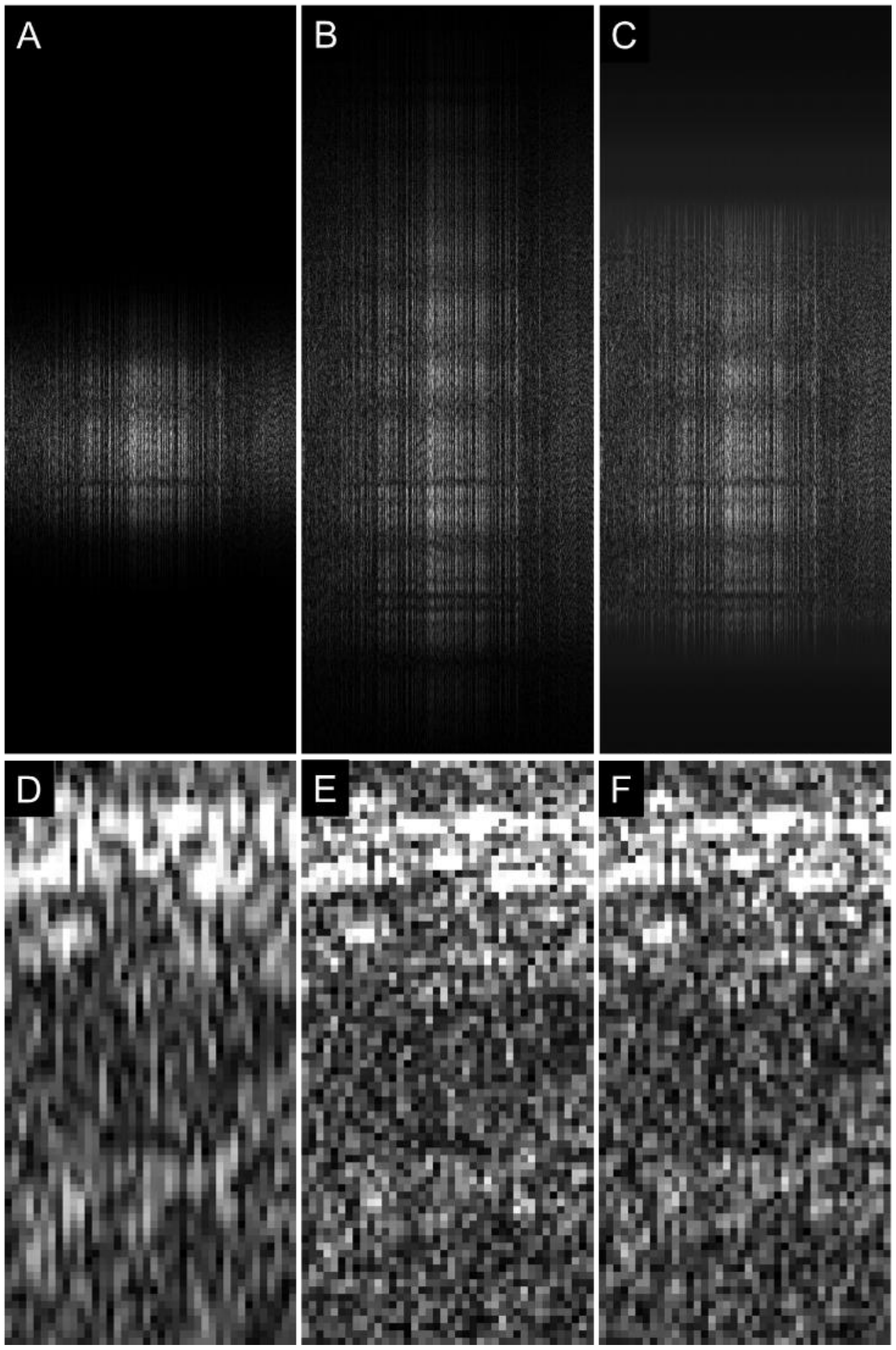

Figure 11. Results of the spectral SRGAN trained on the spectral fringes of OCT B-scan images on the full dataset (35 eyes). Samples from the generated images from the test set. Comparison between (A) windowed fringe data, (B) ground truth fringe data, (C) generated fringe data, (D) windowed spatial domain data, (E) spatial domain data which is Fourier-transformed from the ground truth spectral data, and (F) spatial domain data which is Fourier-transformed from the generated spectral data.

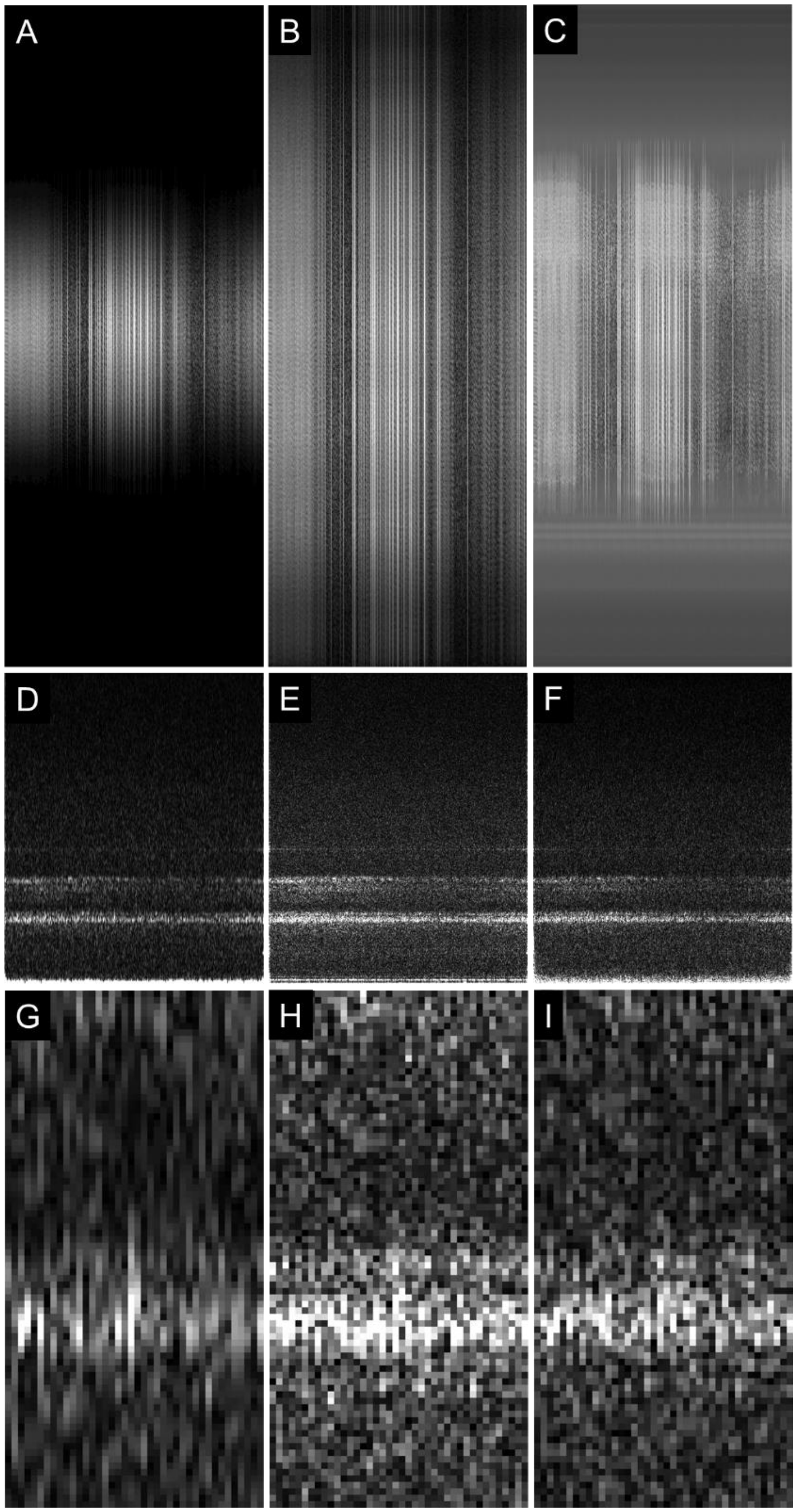

Figure 12. Potential failure case of the SRGAN trained on the spectral fringes of OCT B-scan images on the full dataset (35 eyes). Samples from the generated images from the test set. Comparison between (A) windowed fringe data, (B) ground truth fringe data, (C) generated fringe data, (D) windowed spatial domain data, (E) ground truth spatial domain data, (F) generated spatial domain data, (G) cropped windowed, (H) cropped ground truth, and (I) cropped generated.

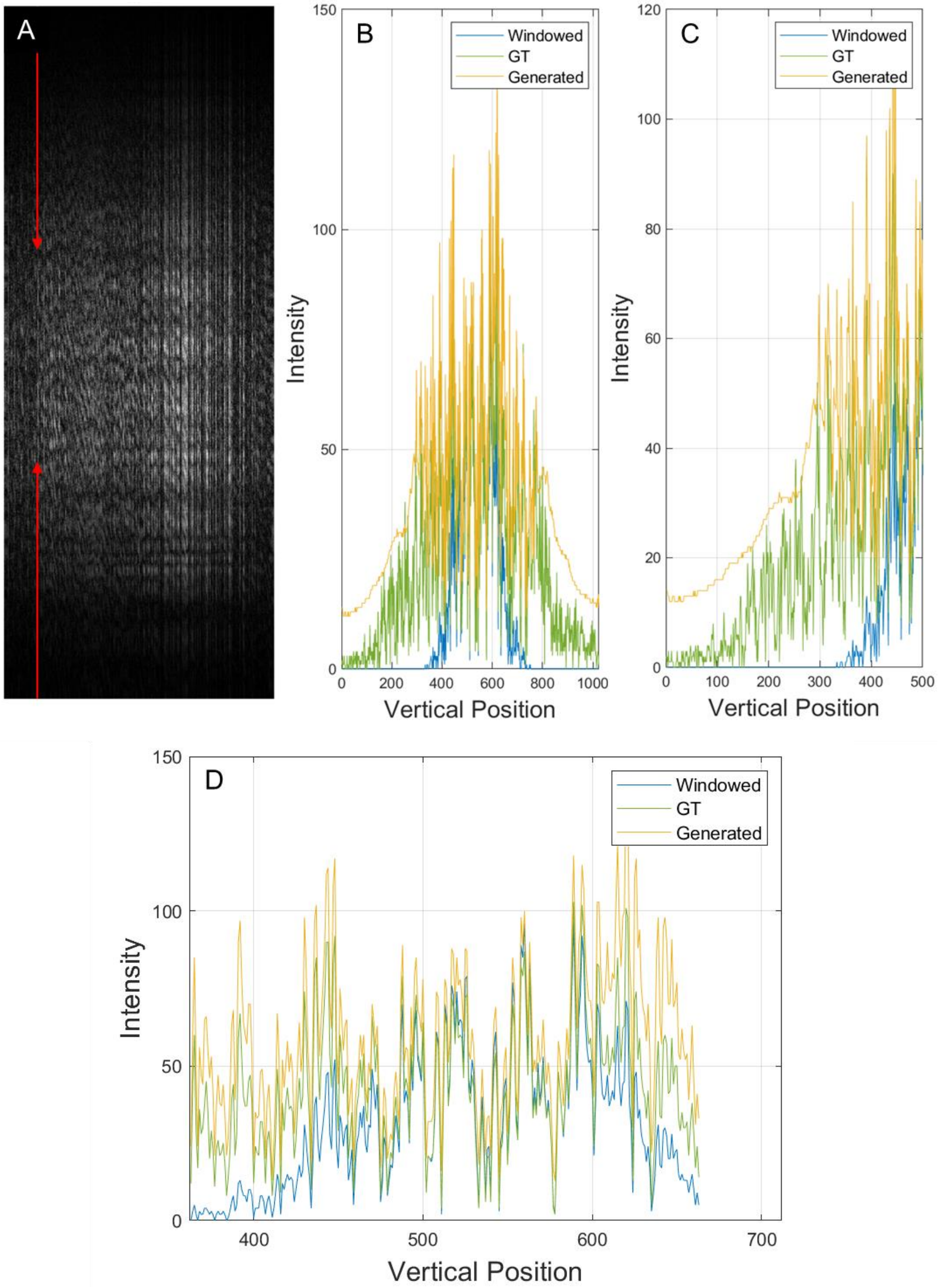

Figure 13. Results of SRGAN trained on the spectral fringes of OCT B-scan images on the full dataset (35 eyes). Samples from the generated images from the test set. The generated A-scans, in yellow, are compared to the ground truth (GT) in orange and windowed fringe in blue. (A) presents the B-scan and the A-scan of interested highlighted by the red arrows. (B) shows the full A-scan in the spectral domain. (C) examines the top half of the image to provide understanding on the generator's performance nearing the edge of the fringes (normally represent high-frequency signal and spectral noise). (D) examines the central 300 pixels.

### *Quantitative Evaluation*

This experiment was performed on the entire dataset and was more comprehensive in the evaluation than the preliminary study. The spatial versus spectral reconstruction was evaluated for PSNR and the MSE was normalized, and square rooted to provide a more standardized range to compare models with different scales. Table 2 quantitatively compares the performance of the reconstruction in the spectral against the spatial domain.

Table 2. Quantitative evaluation of the generative adversarial network (GAN) images comparing reconstructing in the spectral domain versus in the spatial domain. The spectral domain images were transformed to the spatial domain before the evaluation. Mean values are shown with standard deviation in parentheses.

| **Test** | **GAN vs. GT** | | | **Windowed vs. GT** | | |
|---|---|---|---|---|---|---|
| | ***NRMSE*** | ***PSNR*** | ***SSIM*** | ***NRMSE*** | ***PSNR*** | ***SSIM*** |
| Spatial Domain Reconstruction | **0.3390 (0.021)** | **22.31 (1.02)** | **0.6867 (0.041)** | 0.4239 (0.016) | 20.36 (0.886) | 0.579 (0.069) |
| Spectral Domain Reconstruction | **0.3403 (0.044)** | **21.25 (1.03)** | **0.7874 (0.048)** | 0.4394 (0.026) | 18.98 (0.773) | 0.652 (0.029) |

**Best value** between generated and windowed for the specific test is bolded. **Abbreviation**: Normalized Root Mean squared error (NRMSE), peak signal-to-noise ratio (PSNR), and structural similarity (SSIM). Ground truth (GT)

## 6. Discussion

As OCT hardware moves towards capturing larger field of view including more peripheral parts of the retina, the axial resolution may be compromised to minimize the increase in acquisition time for patient comfort and reduced motion artefacts. The resulting reduced bandwidth in the spectral domain hinders the micrometer-resolution which is one of the benefits of utilizing this modality. This study simulates the reduced axial resolution and aims to reconstruct lost features. The contributions of this study are as follows: (1) the effect of transfer learning from a natural dataset for initializing the OCT-SRGAN; and (2) the comparison between reconstructing in the spectral versus spatial domain.

Published studies have investigated the use of an SRGAN on OCT data. GANs have been leveraged for super-resolution through reconstructing features lost through downsampling in the spatial domain [20], speckle removal [21–23], domain adaptation [24], [25], and synthesizing retinal diseases [26] or other imaging modalities [27] among other applications. Recently, groups have begun utilizing GANs for super-resolution. They have simulated low axial resolution OCT data by windowing in the spectral domain and reconstructed in the spatial domain [28–31]. To the best of our knowledge, this study is the first to leverage a GAN to reconstruct a simulated reduced spectral bandwidth in the spectral domain for ophthalmic OCT data.

The preliminary step of this experiment was performed to understand the effect of transfer learning on a GAN-based reconstruction of OCT data. As shown in Figure 9, the SRGAN leveraging the pre-trained weights on the flower dataset successfully reconstructed the OCT data. The features were sharpened especially in the speckle texture of the retinal layers. However, when compared to the evident effect of the GAN on reconstructing the smoothened features on the flower dataset, the effect is minor (comparing Figure 9 C and D). The quantitative evaluation mirrored this sentiment. . By leveraging transfer learning to initialize the SRGAN on natural flower images, the generated images are slightly better in MSE but slightly worse in SSIM than the OCT-only approach. However, all metrics are well within one standard deviation between the two approaches. These results were promising and provided us the confidence to move forward with the entire dataset.

As the preliminary trial progressed, different windowing alpha values were utilized when reducing the spectral bandwidth. The alpha value of 8 was sufficient to visualize the effect of reduced axial resolution. Transfer learning from the floral dataset was relatively successful and the flower-initialized neural network was able to slightly reduce training time. However, the flower dataset was convolved by a mean filter to simulate the effects of reduced axial resolution in the spatial OCT images. For a fair comparison, the spectral domain neural network must also transfer knowledge from a similar domain. The differences in the two approaches supported an approach without transfer learning. The trials that leveraged transfer learning had superior mean values of the evaluation metrics. However, it also resulted in a higher standard deviation for both MSE and SSIM. We decided to approach the second experiment without transfer learning due to the lower variance of the performance and the flexibility it provided. If transfer learning was utilized, each alteration to the neural network would require the same changes to be retrained on the flower dataset.

The floral dataset was selected as proposed in the Literature [8] for reconstructing reduced resolution in radiographs. The complex textures and patterns in the flower petals provided high contrast edges. When convolving the image with a vertical filter, the effect of blurring is more evident than one performed on a more homogenous image. Further exploration into transfer learning from a more similar domain, such as ultrasound, should be performed. This will allow us to understand the impact of transfer learning from a similar domain compared to a dataset selected to exaggerate the desired effect.

The secondary experiment compared reconstruction in the spatial and spectral domain. Both experiments allocated the same eyes for training, validation, and testing. The preprocessing and evaluation were performed using the same functions. However, differences in rounding, saving formats, and the order of the pipeline resulted in minor changes to the images, as shown in Table 2. The column setting the baseline of evaluation metrics comparing the windowed and ground truth images ideally would be identical since the GAN has no impact on either set of images. However, the data was saved as an image for training at different stages in the spectral and spatial domains. Training also required standardization to [-1, 1] before training and [0, 1] for evaluation. All of the differences in the order of standardization, rounding, and processing must be considered when comparing the two domains. To replicate the changes that occur to the generated images, the GT and windowed fringes were also subject to the same processing pipeline.

When evaluating the reconstruction qualitatively, both approaches were successful in reconstructing lost features. In the spatial domain, as shown in Figure 10, the images appear sharper in both the speckle pattern in the background and the retinal layers. However, slight intensity changes are visible even after intensity normalization. When referencing Figure 10 (D-F), a vertical line in the choroidal region (highlighted by the red circle) is approximately 3 pixels long in the ground truth image. In the corresponding windowed and generated images, the feature appears to be 4-5 pixels and 3-4 pixels long, respectively. In the spectral domain, as shown in Figure 11, the generated fringe data (C) is capable of reconstructing features towards the tails of the Gaussian window. Figure 13 graphically shows a randomly selected A-scan highlighted by the arrows in (A). The zoomed-in view of the center of an edge of the A-scan (C) and the center of the A-scan (D) confirms that the generator can reconstruct the features lost from Gaussian windowing. Near the center presented in (D), the generated signal is capable of reconstructing patterns in the ground truth. By examining (C), which is the top half of the signal, the signals appear to be similar up until 200 pixels from the center whereas the windowed signal has nearly reached zero intensity. Thus, we conclude that the minor signals remaining in the windowed image paired with the patterns found near the center of the image are sufficient for the GAN to reconstruct features that have been suppressed through Gaussian windowing.

When evaluating the mean of the metrics, shown in Table 2, the spatial GAN outperforms the spectral GAN in both NRMSE and PSNR and the spectral GAN is superior in SSIM. However, the trends are evident in the right column comparing the windowed and ground truth images. The differences between the two approaches when evaluating the generated images are also within one standard deviation. Thus, we conclude that both approaches are comparable and effective for reconstructing features lost from the reduced spectral bandwidth.

The spectral domain models were optimized within 15 epochs or approximately 100,000 training iterations. A failure case can be seen in Figure 14 demonstrating the effect of overfitting at 19 epochs. The fringe data contained cyclical vertical streaking patterns upon inspection and when converted into the spatial domain, regions above the retina were removed. Conversely, the spatial GANs were able to train up to approximately 80 epochs without overfitting. This issue should be further explored as it could be a result of the similarities between fringe data or the necessity of revising the learning rate scheduler.

## 7. Limitations and Future Works

The SRGAN study was limited by the lack of comparison to a traditional 'deconvolution' method. We have compared the performance of a model that leveraged transfer learning from a natural floral dataset, but lack a fair comparison to other conventional programming reconstruction techniques. The study would also benefit from exploring reconstruction in the spectral domain and investigating other methods of qualitative evaluation. Some groups have adapted a qualitative evaluation into a quantitative evaluation through a mean opinion score, where a group of experts are randomly polled to select the best between the generated, windowed, and other alternative methods. This is a potential avenue of evaluating the benefit and performance of the generated super-resolution images. However, this requires the opinion of experts, and the process may be time-consuming. Another method of evaluation, inspired by my previous work [24], could be through other processing tools by biasing a layer segmentation tool towards the high-resolution domain and evaluating the generated super-resolution image by the performance of the segmentation.

A potential failure case was examined in Figure 12 where the fringe data contained vertical line artefacts. The SRGAN was still able to reconstruct the features despite the visually unappealing lines. Upon visual inspection, the bandwidth did not seem to increase as drastically as other examples. However, when we transform the image into the spatial domain, it is evident that the GAN was still successful in reconstructing the lost features.

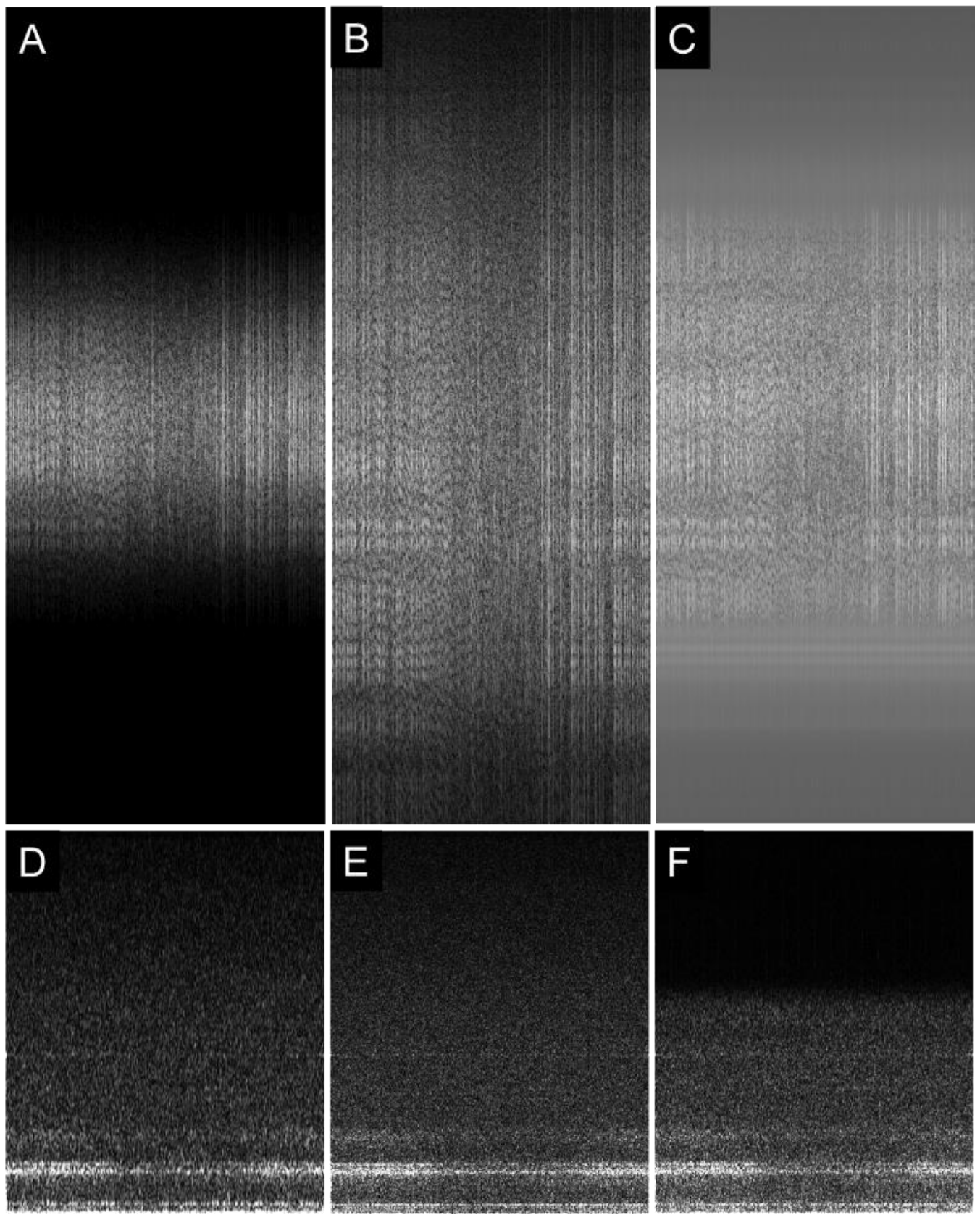

Figure 14. Results of an overfit SRGAN (19 epochs) trained on the spectral fringes of OCT B-scan images on the full dataset (35 eyes). Samples from the generated images from the test set. Comparison between (A) windowed fringe data, (B) ground truth fringe data, (C) generated fringe data, (D) windowed spatial domain data, (E) ground truth spatial domain data, and (F) generated spatial domain data.

Future works includes incorporating the spatial domain A-scans as part of the loss function in the spectral domain GAN, and vice versa. Additional future works include combining GANs from both domains into one processing pipeline, exploring transfer learning from a similar domain of data such as ultrasound or frequency signals from music, and investigating the use of the ResUNet-a architecture on the spatial domain.

## 8. Conflict of Interest

MVS: Seymour Vision, Inc. (I).